\documentclass{svmult}

\usepackage{makeidx}     
\usepackage{multicol}    

\usepackage{amsfonts}
\usepackage{amsmath}

\makeindex             

\def\Zint{\mathbb{Z}}

\def\Real{\mathbb{R}}
\def\Qrat{\mathbb{Q}}
\def\Adele{\mathbb{A}}
\def\href#1{}
\catcode`\@=11
\def\adots{\mathinner{\mkern2mu\raise\p@\hbox{.}
\mkern2mu\raise4\p@\hbox{.}\mkern1mu
\raise7\p@\vbox{\kern7\p@\hbox{.}}\mkern1mu}}
\catcode`\@=12

\begin{document}

\title*{Automorphic forms: a physicist's survey}
\author{Boris Pioline\inst{1}\and Andrew Waldron\inst{2}}
\institute{LPTHE, Universit\'es Paris VI et VII, 4 pl Jussieu, \\
75252 Paris cedex 05, France,
\texttt{pioline@lpthe.jussieu.fr}
\and
Department of Mathematics, One Shields Avenue,\\
University of California, Davis, CA 95616, USA,
 \texttt{wally@math.ucdavis.edu}}
%
%
\maketitle
\begin{abstract}
Motivated by issues in string theory and M-theory, we provide a
pedestrian introduction to automorphic forms and theta series,
emphasizing examples rather than generality. 
\end{abstract}
\setcounter{minitocdepth}{2}
\dominitoc 

\newpage
\index{Automorphic forms}
\index{M-theory}
\index{Superstring dualities}
Automorphic forms play an important r\^ole in physics,
especially in the context of string and M-theory dualities. 
Notably, U-dualities, first discovered
as symmetries of classical toroidal compactifications of  
11-dimensional supergravity by
Cremmer and Julia~\cite{pioline:hidden} and later on elevated to 
quantum postulates by Hull and Townsend~\cite{pioline:Hull:1994ys},
motivate the study of automorphic forms
for exceptional arithmetic groups $E_n(\Zint)$ 
($n=6,7,8$, or their $A_n$ and $D_n$ analogues for $1\leq n\leq5$) --
see {\it e.g.}~\cite{pioline:Obers:1998fb} for a 
review of U-duality.\index{U-duality}
These notes are a pedestrian introduction to these (seemingly 
abstract) mathematical objects, designed to offer a concrete
footing for physicists\footnote{The more mathematically minded reader
may consult the excellent review~\cite{pioline:Miller}.}. 
The basic concepts are introduced via the simple
$Sl(2)$ Eisenstein and theta series. The general
construction of continuous representations and of 
their accompanying Eisenstein series is detailed for~$Sl(3)$. 
Thereafter we present unipotent
representations and their theta series for arbitrary simply-laced groups,
based on our recent work with D. Kazhdan~\cite{pioline:kpw}. We include
a (possibly new) geometrical interpretation of minimal representations, 
as actions on pure spinors or generalizations thereof.
We close with some comments about the
physical applications of automorphic 
forms which motivated our 
research.

\section{Eisenstein and Jacobi Theta series disembodied}

The general mechanism underlying automorphic forms  
is best illustrated by taking a representation-theoretic tour
of two familiar $Sl(2,\Zint)$ examples:

\subsection{$Sl(2,\Zint)$ Eisenstein series}

Our first example is the non-holomorphic Eisenstein series  \index{Eisenstein series!non-holomorphic} \index{Maass waveform|see{Eisenstein series}}
\index{${\cal E}_{s}^{Sl(2)}(\tau)$}
\begin{equation}
\label{pioline:eis2}
{\cal E}^{Sl(2)}_{s}(\tau)=\sum_{(m,n)\in \Zint^2\backslash(0,0)} 
\left(
\frac{\tau_2}{|m+n \tau|^2}
\right)^s\, ,
\end{equation}
which, for $s=3/2$, appears in string theory as the description of 
the complete, non-perturbative, four-graviton
scattering amplitude at low energies~\cite{pioline:Green:1997tv}.
It is a function of the complex modulus $\tau$, taking values on
the Poincar\'e upper half plane\index{Poincar\'e upper half plane}, 
or equivalently points in the
symmetric space 
${\cal M}=K\backslash G =
SO(2)\backslash Sl(2,\Real)$
with coset representative
\begin{equation}
e=\frac{1}{\sqrt{\tau_2}}
\begin{pmatrix} 1 & \tau_1 \\ 0 & \tau_2\end{pmatrix}\, \in Sl(2,\Real).
\end{equation}
The Eisenstein series~(\ref{pioline:eis2}) is
invariant under the modular
transformation 
\begin{equation}
\tau\to(a\tau+b)/(c\tau+d)\, ,
\end{equation}  
which is the right action of
$g\in Sl(2,\Zint)$ \index{$Sl(2,\mathbb{Z})$} on ${\cal M}$. 
Invariance follows simply from that of the lattice $\Zint\times\Zint$. 
This set-up may be formalized 
by introducing: \index{Unitary representation}
\begin{enumerate}
\item[(i)] The linear representation $\rho$ of
$Sl(2,\Real)$ in the space ${\cal H}$ of functions of 
two variables $f(x,y)$,
\begin{equation}
\label{pioline:rho} 
[\rho(g)\cdot f](x,y)=f(ax+by,cx+dy)\ ,\quad
g=\begin{pmatrix} a & b \\ c & d \end{pmatrix}\ ,\quad ad-bc=1\ .
\end{equation}
\item[(ii)]
An  $Sl(2,\Zint)$-invariant
distribution 
\begin{equation}
\label{pioline:deltaz}
\delta_\Zint(x,y)=\sum_{(m,n)\in \Zint^2\backslash(0,0)} 
\delta(x-m)\delta(y-n)
\end{equation} 
in the dual space ${\cal H}^*$. 
\item[(iii)] A vector 
\begin{equation}
\label{pioline:fk}
f_{K}(x,y)=(x^2+y^2)^{-s}
\end{equation} invariant under
the maximal compact subgroup $K=SO(2)\subset G=Sl(2,\Real)$.
\index{Maximal sompact subgroup}
\end{enumerate}
The Eisenstein series~(\ref{pioline:eis2}) may now be 
recast in a general notation for automorphic forms
\begin{equation}
\label{pioline:eisd}
{\cal E}^{Sl(2)}_{s}(e)
=\langle \delta_\Zint , \rho(e)\cdot f_{K} \rangle\ ,\qquad
e\in G\, .
\end{equation}
The modular invariance of ${\cal E}^{Sl(2)}_{s}$ 
is now manifest: 
under the right action $e$~$\to$~$eg$ of $g\in Sl(2,\Zint)$, the vector 
$\rho(e)\cdot f_K$ transforms by $\rho(g)$, which in turn hits 
the $Sl(2,\Zint)$   invariant distribution $\delta_\Zint$.
Furthermore~(\ref{pioline:eisd}) is ensured to be
a function of the {\it coset} $K\backslash G$ 
by invariance of the vector $f_{K}$ under the maximal compact
$K$. Such a distinguished vector is  known as
{\it spherical}. \index{Spherical vector} All the automorphic forms 
we shall encounter can be written in terms of a triplet 
$(\rho,\delta_\Zint, f_{K})$.

Clearly any other function of the $SO(2)$ invariant 
norm $|x,y|_\infty\equiv\sqrt{x^2+y^2}$ would
be as good a candidate for~$f_{K}$. 
This reflects the reducibility of
the representation~$\rho$ in~(\ref{pioline:rho}).
However, its restriction to homogeneous, even functions of
degree~$2s$,
\begin{equation}
\label{pioline:homog}
f(x,y)=\lambda^{2s} \, f(\lambda x,\lambda y) = y^{-2s} f\Big(\frac
xy,1\Big)
\, ,
\end{equation}
is irreducible. The restriction of the representation $\rho$  acts
on the space of functions of a single variable $z=x/y$ by weight $2s$
conformal
transformations $z\to (az+b)/(cz+d)$ and admits 
$f_K(z)=(1+z^2)^{-s}$ as its unique spherical vector. 
In these variables, the distribution
$\delta_\Zint$ is rather singular as its support is on all
rational values $z\in \Qrat$. A related problem is that the
behavior of ${\cal E}^{Sl(2)}_{s}(\tau)$ 
at the cusp $\tau\to i \infty$ is difficult
to assess -- yet of considerable interest to physicists being 
the limit relevant to non-perturbative instantons~\cite{pioline:Green:1997tv}.

These two problems may be evaded by performing a Poisson resummation
on the integer $m\rightarrow\widetilde m$ in the sum (\ref{pioline:deltaz}),
after first separating out terms with $n=0$. 
The result may be rewritten
as a sum over the single variable $N=\widetilde m n$, except for two
degenerate -- or {\it ``perturbative''} -- contributions:
\begin{align}
\label{pioline:inst}
{\cal E}^{Sl(2)}_{s} &= 2\ \zeta(2s)\ \tau_2^{s}
+ \frac{2 \sqrt{\pi}\ \tau_2^{1-s}\ \Gamma(s-1/2)\ \zeta(2s-1)}{\Gamma(s)}
\nonumber\\
&+\frac{2\pi ^s\sqrt{\tau_2}}{\Gamma(s)}
\sum_{N\in\Zint\backslash\{0\}} \mu_{s}(N)\; N^{s-1/2} 
K_{s-1/2} \left( 2 \pi \tau_2N \right) 
e^{2\pi i  \tau_1N} \, .
\end{align}
In this expression, the summation measure \index{Divisor function}  \index{$\mu_{s}(N)$}
\begin{equation}\mu_s(N)=\sum_{n| N} n^{-2s+1}\, ,
\label{pioline:measure}\end{equation}
is of prime physical interest, as it is connected to quantum 
fluctuations in an instanton background~\cite{pioline:Gutperle,
pioline:Moore,pioline:Sugino}.

First focus on the non-degenerate terms in the second line.
Analyzing the transformation properties under the 
Borel and Cartan $Sl(2)$ generators 
$\rho\begin{pmatrix} 1 & t \\ 0 & 1 \end{pmatrix}:\tau_1 \to \tau_1+t$ 
and 
$\rho\begin{pmatrix} t^{-1} & 0 \\ 0 & t \end{pmatrix}: 
\tau_2 \to t^2\tau_2$, 
we readily see that they fit into the framework~(\ref{pioline:eisd}),
upon identifying
\begin{equation}
\label{pioline:de66}
f_{K}(z)=z^{s-1/2} K_{s-1/2}(z)\ ,\quad
\delta_\Zint(z)=\sum_{N\in\Zint\backslash\{0\}} \mu_{s}(N)\ \delta(z- N)\, ,
\end{equation}
and the representation $\rho$ as
\begin{equation}
\label{pioline:e13}
E_+=iz\ , \quad 
E_-=i(z\partial_z +2-2s) \partial_z ,\quad
H=2z\partial_z+2-2s\, .
\end{equation}
This is of course equivalent to the representation on
homogeneous functions~(\ref{pioline:homog}),
upon Fourier transform in the variable $z$. 
The power-like degenerate terms in~(\ref{pioline:inst}) may be 
viewed as regulating the singular value of the 
distribution $\delta$ at $z=0$.
They may, in principle, be recovered by performing a Weyl reflection on the
regular part.
It is also easy to check that the spherical vector condition,
$K\cdot f_K(z)\equiv (E_+-E_-)\cdot f_K(z)=0$, 
is the modified Bessel equation whose unique
decaying solution at $z\to \infty$ is the spherical vector in~(\ref{pioline:de66}). 

While the representation $\rho$ and its
spherical vector $f_{K}$ are easily understood,
the distribution $\delta_\Zint$ requires additional technology. 
Remarkably, the summation measure~(\ref{pioline:measure})
can be written as an infinite product
\begin{equation}
\label{pioline:mup}
\mu_{s}(z)=\prod_{p{\rm\;  prime}} f_p(z)\ ,
\qquad
f_p(z) = \frac{1-p^{-2s+1}|z|_p^{2s-1}}{1-p^{-2s+1}} \gamma_p(z)\, .
\end{equation}
(A simple trial computation of $\mu_s(2\cdot 3^2)$ will easily 
convince the reader of this equality.)
Here $|z|_p$ is the $p$-adic\footnote{A useful
physics introduction to $p$-adic and adelic fields
is~\cite{pioline:Brekke:1993}\index{Adelic}. It is worth noting that a special function
theory analogous to that over the complex numbers exists for
the $p$-adics.}
norm of $z$, {\it i.e.} $|z|_p=p^{-k}$ with $k$ 
the largest integer such that $p^k$ divides $z$. The function
$\gamma_p(z)$ is unity if $z$ is a $p$-adic integer ($|z|_p\leq 1$) and 
vanishes otherwise.
Therefore $\mu(z)$ vanishes unless $z$ is an integer $N$.
Equation~(\ref{pioline:eisd}) can therefore be expressed 
as \begin{equation}
\label{pioline:adelsum}
{\cal E}_{s}^{Sl(2)}(e)= \sum_{z\in \Qrat} 
\prod_{p\ =\ {\rm prime},\infty}f_p(z) 
\rho(e) \cdot f_K(z)\ ,
\end{equation}
The key observation now is that $f_p$ 
is in fact the spherical vector for the 
representation of $Sl(2,\Qrat_p)$, just as  
$f_\infty:= f_K$ is the spherical vector
of $Sl(2,\Real)$~! In order to convince herself of this
important fact, the reader may evaluate the $p$-adic Fourier transform
of $f_p(y)$ on $y$, 
thereby reverting to the $Sl(2)$ representation 
on homogeneous functions~(\ref{pioline:homog}): the result 
\begin{equation}
\tilde f_p(x) = \int_{\Qrat_p} dz \,f_p(z) e^{i x z} =  | 1 , x
|_p^{-2s}
\equiv \max(1,|x|_p)^{-2s},
\end{equation}
is precisely the $p$-adic counterpart of the real spherical vector 
$f_K(x)=(1+x^2)^{-s}\equiv |1,x|_{\infty}^{-2s}$. The analogue
of the decay condition is that $f_p$ should have support over
the $p$-adic integers only, which holds by virtue of the 
factor $\gamma_p(y)$ in~(\ref{pioline:mup}).
It is easy to check that the
formula~(\ref{pioline:adelsum}) in this representation 
reproduces the Eisenstein series~(\ref{pioline:eis2}).

Thus, the $Sl(2,\Zint)$-invariant distribution $\delta_\Zint$
can be straightforwardly obtained by computing the spherical
vector over all $p$-adic fields $\Qrat_p$. More conceptually,
the Eisenstein series~(\ref{pioline:eis2}) may be written {\it adelically}
\index{Adelic} 
(or {\it globally}) as \index{Eisenstein series!adelic representation}
\begin{equation}
\label{pioline:adelsum2}
{\cal E}_{s}^{Sl(2)}(e)= \sum_{z\in \Qrat} \rho(e) \cdot f_{\Adele}(z)
\ ,\quad f_{\Adele}(z)=\prod_{p\ =\ {\rm prime},\infty}f_p(z) \, ,
\end{equation}
where the sum $z\in \Qrat$ is over principle 
adeles\footnote{Adeles are infinite sequences $(z_p)_{p={\rm prime},\infty}$
where all but a finite set of $z_p$
are $p$-adic integers. Principle adeles are constant sequences $z_p=z
\in \Qrat$, isomorphic to $\Qrat$ itself.}, and 
$f_{\Adele}$ is the spherical vector of $Sl(2,\Adele)$,
invariant under the maximal compact subgroup $K(\Adele)=
\prod_p Sl(2,\Zint_p) \times U(1)$ of $Sl(2,\Adele)$.
This relation between functions on 
$G(\Zint)\backslash$ $G(\Real)/\!\!$ $K(\Real)$
and functions on $G(\Qrat)\backslash G(\Adele)/K(\Adele)$ 
is known as the Strong
Approximation Theorem\index{Strong Approximation Theorem}, 
and is a powerful tool
in the study of automorphic forms (see e.g.~\cite{pioline:Miller}
for a more detailed introduction to the adelic approach).

\subsection{Jacobi theta series} \index{Theta series!Jacobi}
Our next example, the Jacobi theta series,
demonstrates the key r\^ole played by Fourier invariant Gaussian
characters -- {\it ``the Fourier transform of the Gaussian is the Gaussian''}. 
Our later generalizations will involve
cubic type characters invariant under Fourier transform.

In contrast to the Eisenstein series, the Jacobi theta series
\begin{equation}
\label{pioline:e1}
\theta(\tau)=\sum_{m\in\Zint} e^{i\pi  \tau m^2}\, ,
\end{equation} 
is a modular form for a congruence subgroup $\Gamma_0(2)$ of $Sl(2,\Zint)$
with  modular weight~$1/2$ and a non-trivial multiplier system. 
It may, nevertheless, be cast in the 
framework~(\ref{pioline:eisd}), with a minor {\it caveat}. 
The representation $\rho$ now 
acts on functions of a single variable $x$ as
\begin{equation}
\label{pioline:e133}
E_+= i\pi \,x^2\, , 
\quad H=\frac{1}{2}\,(x\partial_x+\partial_x x)\, ,\quad
E_-= \frac{i}{4\pi} \partial_x^2\, ,
\end{equation}
Here, the action of $E_+$ and $H$ may be read off from the usual 
Borel and Cartan actions of $Sl(2)$ on $\tau$ while
the generator $E_-$ follows by noting that the Weyl reflection
$S:\tau\to -1/\tau$ can be compensated 
by Fourier transform on the integer $m$. The invariance of 
the ``comb'' distribution $\delta_{\Zint}(x)=\sum_{m\in\Zint} \delta(x-m)$
under Fourier transform is just the Poisson resummation formula.

Finally (the {\it caveat}), the compact generator $K=E_+ - E_-$ is exactly the
Hamiltonian of the harmonic oscillator, which notoriously does {\it not} 
admit a normalizable zero energy eigenstate, but rather the 
Fourier-invariant ground state $f_{\infty}(x)=e^{-\pi x^2}$ 
of eigenvalue $i/2$. 
This relaxation of the spherical vector condition is responsible for
the non-trivial modular weight and multiplier system.
Correspondingly, $\rho$ does not represent the
group $Sl(2,\Real)$, but rather its double cover, the metaplectic group.
\index{Unitary representation!Metaplectic}

Just as for the Eisenstein series, an adelic formula for the
summation measure exists: note that the $p$-adic spherical vector
must be invariant under the compact generator $S$ which acts by Fourier
transform. Remarkably, the function $f_p(x)=\gamma_p(x)$, imposing
support on the integers only is Fourier invariant -- it is the
$p$-adic Gaussian!\index{$\gamma_p(x)$}\index{Gaussian} 
One therefore recovers the ``comb'' distribution with uniform measure.
Note that  the $Sl(2)=Sp(1)$ theta series generalizes 
to higher symplectic groups
under the title of Siegel theta series\index{Theta series!Siegel}, 
relying in the same way on
Gaussian Poisson resummation.

\section{Continuous representations and Eisenstein series}

\index{Unitary representation!Continuous}
The two $Sl(2)$ examples demonstrate that 
the essential ingredients for automorphic forms with respect to
an arithmetic group $G(\Zint)$ are (i) an irreducible representation $\rho$
of $G$ and (ii) corresponding 
spherical vectors over $\Real$ and $\Qrat_p$. We
now explain how to construct 
these representations by quantizing  coadjoint orbits.

\subsection{Coadjoint orbits, classical and quantum: $Sl(2)$}
As emphasized by Kirillov, unitary representations are quite
generally in correspondence with coadjoint orbits~\cite{pioline:Kirillov}.
\index{Orbit method}
For simplicity, we restrict ourselves to finite, simple, Lie algebras 
${\mathfrak g}$, where the Killing form $(\cdot,\cdot)$ 
identifies ${\mathfrak g}$ with
its dual. Let ${\cal O}_j$ be the orbit of an element $j\in {\mathfrak g}$
under the action of $G$ by the adjoint representation 
$j\to g j g^{-1}\equiv\tilde j$.
Equivalently, ${\cal O}_j$ may be viewed as an homogeneous space 
$S\backslash G$, where $S$ is the stabilizer (or commutant) of $j$. 

The (co)adjoint orbit ${\cal O}_j$ admits a (canonical, up to
a multiplicative constant) $G$-invariant Kirillov--Kostant
symplectic form, \index{Kirillov--Kostant symplectic form} defined on
the tangent space at a point $\tilde j$ on the orbit
by $\omega(x,y)=(\tilde j,[x,y])$. Non-degeneracy of $\omega$
is manifest, since its kernel,
the commutant $\tilde S$ of $\tilde j$, is gauged away in the 
quotient $S\backslash G$. Parameterizing ${\cal O}_j$
by an element $e$ of $S\backslash G$, one may rewrite
$\omega=d\theta$ where the ``contact'' one-form 
$\theta=(j,de\ \! e^{-1})$, making the closedness
and $G$-invariance of $\omega$ manifest. The coadjoint orbit 
${\cal O}_j=S\backslash G$
therefore yields a classical phase space with a $G$-invariant 
Poisson bracket and hence a set of canonical generators 
representing the action of $G$ on functions of ${\cal O}_j$.
The representation $\rho$ associated to $j$ follows by quantizing
this classical action, {\it i.e.} by choosing a Lagrangian subspace
${\cal L}$ (a maximal commuting set of observables) and representing
the generators of $G$ as suitable differential operators on functions on
${\cal L}$.

This apparently abstract construction is simply illustrated for
$Sl(2)$: consider the coadjoint orbit 
of the element 
\begin{equation}
j=\begin{pmatrix}\frac l2&\\&-\frac l2\end{pmatrix}\, ,
\end{equation}
with stabilizer $S=\Real j$. 
The quotient $S\backslash G$ may be parameterized as
\begin{equation}
\label{pioline:e14}
e=\begin{pmatrix}
1&\\\gamma&1
\end{pmatrix}
\begin{pmatrix}
1&\beta\\&1
\end{pmatrix}\, .
\end{equation}
The contact one-form is 
\begin{equation}
\theta={\rm tr}\ j\ \! de \ \!  e^{-1}=-l \gamma d\beta\, .
\end{equation}
The group $G$ acts by right 
multiplication on $e$, followed by a compensating left multiplication by $S$ 
maintaining the choice of gauge slice~(\ref{pioline:e14}). 
The resulting infinitesimal group action is expressed in  terms of
Hamiltonian vector fields 
\begin{equation}
E_+=i\partial_\beta, \quad H=2i \beta \partial_\beta-
2\gamma \partial_\gamma, \quad
E_-=-i\beta^2 \partial_\beta+i(1+2\beta\gamma) \partial_\gamma\, .
\label{pioline:hamvecs}
\end{equation}
We wish to express these transformations in terms of the Poisson bracket 
determined by the Kirillov--Kostant symplectic form
\begin{equation}
\omega=d\theta=l\ d\gamma\wedge d\beta\, ,
\end{equation}
namely
\begin{equation}
\{\gamma,\beta\}_{PB}=\frac1{l}\, .
\label{pioline:PB}
\end{equation}
Indeed, it is easily verified that the generators~(\ref{pioline:hamvecs}) can
be represented canonically 
\begin{equation}
E_+ = il \gamma\ ,\quad
H=2il \beta \gamma \ ,\quad
E_- = -il \beta ( 1 + \beta \gamma)\, ,
\label{pioline:can_gens}
\end{equation}
with respect to  the Poisson bracket~(\ref{pioline:PB}).
The next step is to quantize this classical mechanical system:
\begin{equation}
\gamma=\frac1l\ y\, ,\qquad \beta=\frac1i\frac{\partial}{\partial y}\, .
\label{pioline:quantize}
\end{equation} 
The quantized coadjoint orbit representation follows directly by 
substituting~(\ref{pioline:quantize}) in~(\ref{pioline:can_gens}) and the result
is  precisely the Eisenstein series
representation~(\ref{pioline:e13}). The physicist reader will observe that
the parameter $s$ appearing there arises from quantum orderings
of the operators $\beta$ and $\gamma$.

The construction just outlined, based on the quantization of an
element $j$ in the {\it hyperbolic} conjugacy class of $Sl(2,\Real)$,
leads to the continuous series representation of $Sl(2,\Real)$. 
Recall that conjugacy classes of $Sl(2)$ are classified by 
the value
\footnote{The geometry 
of the
three coadjoint orbits is exhibited by 
parameterizing the~$sl(2)$ Lie algebra as
$
{\mathfrak g}=
\begin{pmatrix}
k_1&k_2+k_0\\k_2-k_0&-k_1
\end{pmatrix}\, .
$
The orbits are then seen to correspond to massive, lightlike and 
tachyonic $2+1$ dimensional mass-shells
$
k_\mu k^\mu=-k_0^2+k_1^2+k_2^2=-\frac{C}{4}\, .
$} of $C\equiv 2\ {\rm tr}j^2=l^2>0$.
The elliptic case $C<0$ with $j$
conjugate to an antisymmetric matrix leads to discrete series
representations and will not interest
us in these Notes. However, the non-generic parabolic (or nilpotent)
\index{Nilpotent orbit}
conjugacy class
$C=0$ is of considerable interest, being key to theta series
for higher groups. 
There is only a single nilpotent conjugacy class
with representative
\begin{equation}
j=\begin{pmatrix}&\\1&\ \ \end{pmatrix}\, ,\qquad j^2=0\, .
\end{equation} 
The stabilizer $S\subset Sl(2,\Real)$ is the parabolic group of
lower triangular matrices so the nilpotent orbit $S\backslash G$
may be parameterized as
\index{Parabolic subgroup}
\begin{equation}
e=\frac{1}{\sqrt \gamma}\begin{pmatrix}
\gamma&\\&1
\end{pmatrix}
\begin{pmatrix}
1&\beta\\&1
\end{pmatrix}\, .
\end{equation}
The contact and symplectic forms are now 
\begin{equation}
\theta=\gamma d\beta\, ,\qquad
\omega=d\gamma\wedge d\beta\, ,
\end{equation}
and
the action of $Sl(2)$ may be  represented by the canonical generators
\begin{equation}
E_+=i\gamma\ ,\quad H=2i\beta \gamma\ ,\quad E_-=-i\beta^2 \gamma
\end{equation}
accompanied by Poisson bracket $\{\gamma,\beta\}_{PB}=1$. This representation
also follows by the contraction $l\rightarrow 0$ holding $l\gamma$
fixed in~(\ref{pioline:can_gens}).
The relation to theta series is
exhibited by performing a canonical transformation 
$\gamma=y^2$ and $\beta=\frac12 p/y$ which yields
\begin{equation}
E_+=iy^2\ ,\quad 
H=ip y\ ,\quad
E_-=-\frac i4 p^2\, .
\label{pioline:metrep}
\end{equation}
Upon quantization, this is 
precisely the metaplectic representation in~(\ref{pioline:e133}).
\index{Unitary representation!Metaplectic} 
In contrast to the continuous series, 
there is no quantum ordering parameter (although  a peculiarity of $Sl(2)$
is that it appears as the $s=1$ instance of the continuous series
representation~(\ref{pioline:e13})).

\subsection{Coadjoint orbits: general case} \index{Coadjoint orbits}

For general groups $G$, the orbit method predicts
the Gelfand-Kirillov dimension\footnote{The Gel'fand--Kirillov, or 
functional dimension counts the
number of variables -- being unitary, all these representations
of non-compact groups are of course infinite dimensional in the usual sense.}
\index{Gel'fand--Kirillov dimension}
of the generic continuous 
irreducible representation to be $(\dim G - {\rm rank}~G)/2$: 
a generic non-compact element may be conjugated into the Cartan
algebra, whose stabilizer is the Cartan (split) torus. There are,
therefore,
${\rm rank}~G$ parameters corresponding to the eigenvalues in the
Cartan subalgebra. Non-generic elements arise when eigenvalues collide,
and lead to representations of smaller functional dimension. When all
eigenvalues degenerate to zero, there are a finite set of
conjugacy class of nilpotent elements with non-trivial Jordan
patterns, hence a finite set of parameter-less
representations usually called ``unipotent''. 
The nilpotent orbit of smallest dimension, namely
the orbit of any root, leads to the {\it minimal}
\index{Unitary representation!Minimal} 
unipotent representation,  which plays a
distinguished r\^ole as the analog of the $Sl(2)$ (Jacobi
theta series) metaplectic representation~\cite{pioline:Joseph}.

\subsection{Quantization by induction: $Sl(3)$} \index{$Sl(3)$}
\index{Unitary induction}

Given a symplectic manifold
with $G$-action, there is no {\it general} 
method to resolve the quantum ordering
ambiguities while maintaining the ${\mathfrak g}$-algebra.
However, (unitary) induction provides a standard procedure
to extend a representation $\rho_H$ of a subgroup $H\subset G$
to the whole of $G$. 
Let us illustrate the first non-trivial case: 
the generic orbit of $Sl(3)$.

Just as for $Sl(2)$ in~(\ref{pioline:e14}), 
the coadjoint orbit of a generic $sl(3)$ Lie algebra element
\begin{equation}
j=
\begin{pmatrix}
l_1&&\\
&l_2&\\
&&l_3
\end{pmatrix}
\, , 
\end{equation}
can be parameterized by
the gauge-fixed $Sl(3)$ group element
\begin{equation}
\label{pioline:sl3g}
e = 
\begin{pmatrix}
1 & & \\
y & 1 & \\
w+yu & u & 1
\end{pmatrix}
\cdot
\begin{pmatrix}
1 & x & v+xz \\
  & 1 & z \\
  &   & 1
\end{pmatrix}\, ,
\end{equation}
whose  six-dimensional phase space is equipped with the 
contact one-form
\begin{equation}
\theta=(l_2-l_1) y dx + (l_3-l_2) u dz + [(l_3-l_1)w+(l_3-l_2)y u]
(dv+ x dz)\, .
\end{equation}
(The canonical generators are easily calculated.) 
To quantize this orbit, a natural choice of 
Lagrangian submanifold is $w=y=u=0$ so that $Sl(3)$  
is realized on functions of three variables $(x,z,v)$. 
These variables parameterize the coset $P\backslash G$,
where $P=P_{1,1,1}$ 
is the (minimal) parabolic subgroup of lower triangular matrices (look
at equation~(\ref{pioline:sl3g})). \index{Parabolic subgroup}
A set of one-dimensional representations on $P$ are realized by
the character
\begin{equation}
\chi(p)=\prod_{i=1}^3 |a_{ii}|^{\rho_i} \mbox{sgn}^{\epsilon_i}(a_{ii})
\ ,\quad
p=\begin{pmatrix}
a_{11} & &\\
a_{21} & a_{22} &\\
a_{31} & a_{32} & a_{33} \end{pmatrix}
\in P
\, ,
\label{pioline:character}
\end{equation}
where $\rho_i$ are three constants (defined up to a common shift
$\rho_i \to \rho_i + \sigma$) and $\epsilon_i\in\{0,1\}$ are three discrete
parameters. The representation of $G$ on functions of $P\backslash G$
induced from $P$ and its character representation~(\ref{pioline:character})
acts by
\begin{equation}
g:f(e)\mapsto \chi(p) f(eg^{-1})\, , 
\end{equation}
where $eg^{-1}=pe'$ and $e'\in P\backslash G$ (coordinatized by $\{x,z,v\}$).
It is straightforward to obtain the corresponding generators explicitly,
\begin{eqnarray}
\begin{array}{lc}
E_{\beta}=\partial_x -z \partial_v&
E_{-\beta}=x^2\partial_x+v\partial_z+(\rho_2-\rho_1)x\\
E_{\gamma}=\partial_z&
E_{-\gamma}=z^2\partial_z+vz\partial_v-(v+xz)\partial_x +(\rho_3-\rho_2)z\\
E_{\omega}=\partial_v&
E_{-\omega}=v^2\partial_v+vz\partial_z+x(v+xz)\partial_x 
+(\rho_3-\rho_1)v+(\rho_2-\rho_1) xz
\end{array}\nonumber
\end{eqnarray}
\begin{equation}
\hspace{-.03mm} H_\beta=2x\partial_x+v\partial_v-z\partial_z+(\rho_2-\rho_1)\,\qquad
H_\gamma=-x\partial_x+v\partial_v+2z\partial_z+(\rho_3-\rho_2)\, ,
\end{equation}
where $Sl(3)$ generators are defined by,
\begin{equation}
sl(3)\ni X = 
\begin{pmatrix} 
-\frac23 H_\beta-\frac13 H_\gamma & E_{\beta} & E_{\omega} \\ 
-E_{-\beta} & -\frac13 H_\gamma+\frac13 H_\beta & E_{\gamma} \\
-E_{-\omega} & -E_{- \gamma} & \frac23 H_\gamma+\frac13 H_\beta
\end{pmatrix}\, .
\end{equation}
For later use, we evaluate the action of the Weyl reflection
$A$ with respect to the root $\beta$ which exchanges 
the first and second rows of $e$ up to a compensating $P$ transformation,
\begin{equation}
[A\cdot f] (x,v,z) = x^{\rho_2-\rho_1} f(-z,v,-1/x)\, .
\end{equation}
The quadratic and cubic Casimir invariants $C_2=\frac12\mbox{Tr}~X^2$
and $C_3=\frac{27}2\det X$, 
\begin{eqnarray}
C_2&=&\phantom{+}\frac16\ \left[ 
(\rho_1-\rho_2)^2+(\rho_2-\rho_3)^2+(\rho_3-\rho_1)^2 \right] 
+ (\rho_1-\rho_3)\, , \\
C_3&=&-\frac12\ 
\left[\rho_1+\rho_2-2\rho_3+3\right]
\left[\rho_2+\rho_3-2\rho_1-3\right]
\left[\rho_3+\rho_1-2\rho_2\right]\, ,
\end{eqnarray}
agree with those of the classical representation on the 6-dimensional
phase space $\{x,y,z,u,v,w\}$, upon identifying $l_i=\rho_i$ 
and removing the subleading ``quantum ordering terms''.

The same procedure works in the case of a nilpotent coadjoint orbit.
As an Exercise, the reader may show that the maximal nilpotent
orbit of a single $3\times 3$ Jordan block has 
dimension~6 and can be quantized by induction from the same minimal
parabolic $P_{1,1,1}$. The nilpotent orbit corresponding to an $2+1$
block decomposition on the other hand has dimension~4, leading
to a unitary, functional dimension~2, representation of $Sl(3)$
induced from the (maximal) parabolic $P_{2,1}$. \index{Parabolic subgroup} 
This is the minimal representation
of $Sl(3)$, or simpler, the $Sl(3)$ action
on functions of projective $\Real P^3$. 

\index{Unitary representation!of $Sl(3,\Real)$} 
In fact, all irreducible unitary representations of 
$Sl(3,\Real)$ are classified as representations induced from
(i) the maximal parabolic subgroup $P_{1,1,1}$ by the character 
$\chi(p)$ (with $\rho_i\in i{\mathbb C}$), or
(ii) the parabolic subgroup $P_{1,2}$
by an irreducible unitary representation of $Sl(2)$ of the
discrete, supplementary or degenerate series~\cite{pioline:vahu}.

\subsection{Spherical vector and Eisenstein series}\index{Spherical vector}
The other main automorphic form ingredient, the spherical vector, turns out
to be straightforwardly computable in the $Sl(n)$
representation unitarily induced from the parabolic
subgroup $P$. We simply need a $P$-covariant, $K$-invariant function
on $G$. For simplicity, consider again $Sl(3)$ and denote
the three rows of the second matrix in~(\ref{pioline:sl3g}) as $e_1,e_2,e_3$.
Under left multiplication by a lower triangular matrix $p=(a_{i\leq j})\in P$, 
$e_1\mapsto a_{11}e_1$ and $e_2\mapsto a_{21} e_1 + a_{22} e_2$. Therefore 
the norms of  $|e_1|_\infty$ and $|e_1 \wedge e_2|_\infty$ 
are $P$-covariant and
maximal compact $K=SO(3)$-invariant. 
The spherical vector over $\Real$ 
is the product of these two
norms raised to powers corresponding to the character $\chi$ in~(\ref{pioline:character}),
\begin{equation}
f_{\infty}=\left|1,x,v+xz\right|_\infty^{\rho_1-\rho_2}~
\left|1,v,z\right|_\infty^{\rho_2-\rho_3}\, .
\end{equation}
(Recall that $|\cdot\cdot|_\infty$ is just the usual orthogonal
Euclidean norm.) 
Similarly, the spherical vector over $\Qrat_p$ is the product
of the $p$-adic norms,
\begin{equation}
f_{p}=|1,x,v+xz|_p^{\rho_1-\rho_2} |1,v,z|_p^{\rho_2-\rho_3}\, . 
\end{equation}
The $Sl(3,\Zint)$, continuous series representation, Eisenstein series
follows by summing over principle
adeles, \index{${\cal E}_{\rho_i}^{Sl(3)}(e)$}
\begin{equation}
{\cal E}_{\rho_i}^{Sl(3)}(e) = 
\sum_{(x,z,v)\in \Qrat^3} \left[ \prod_{p~{\rm prime}} f_p
\right] \rho(e) \cdot  f_{\infty}\, .
\end{equation}
Writing out the adelic product in more mundane terms, 
\begin{equation}
{\cal E}_{\rho_i}^{Sl(3)}(e) = 
\sum_{\substack{ (m^i,n^i)\in\Zint^6, \\ m^{ij} \neq 0}}
\left[ (m^{ij})^2 
\right]^\frac{\rho_1-\rho_2}{2}~
\left[ (m^i)^2 \right]^\frac{\rho_2-\rho_3}{2}\,  ,
\end{equation}
where $m^{ij}=m^in^j-m^jn^i$.
As usual, the sum is convergent for ${\rm Re}(\rho_i-\rho_j)$
sufficiently large and can be
analytically continued to complex $\rho_i$ using functional relations
representing the Weyl reflections on the weights $(\rho_i)$.
The above procedure  suffices to describe Eisenstein
series for all finite Lie groups.

\subsection{Close encounters of the cube kind}
\label{pioline:gafgg}

Cubic characters are central to the construction of minimal
representations and their theta functions for higher
simply laced groups $D_n$ and $E_{6,7,8}$.
They can also be found 
in a particular realization of the $Sl(3)$ continuous series representation
at $\rho_i=0$ (which also turns out to arise by restriction 
of the minimal representation of $G_2$~\cite{pioline:Joseph}) :
let us perform the following (mysterious)
sequence of transformations: (i)~Fourier transform over $v,z$, and call
the conjugate variables $\partial_z=ix_0, \partial_v=iy$. 
(ii)~Redefine $x=1/(p y^2) + x_0/y$. 
(iii)~Fourier transform over $p$ and redefine the
conjugate 
variable\footnote{This sequence of transformations also makes sense
at $\rho_2\neq \rho_3$ as long as $\rho_1=\rho_2$.} $p_p=x_1^3$.
These operations yield generators,
\begin{equation}
\begin{array}{lc}
E_{\beta}=y \partial_0&
E_{-\beta}= - x_0 \partial + i \frac{x_1^3}{y^2}\\ \\
E_{\gamma}=i x_0&
E_{-\gamma}= - i(y\partial+x_0 \partial_0 +x_1 \partial_1)\partial_0\\
&\qquad\qquad\quad 
+\frac{1}{27} y \partial_1^3+\frac{4 y \partial_1^2}{9 x_1}
+\frac{28 y \partial_1}{27 x_1^2} -6i \partial_0 \\ \\
E_{\omega}=i y&
E_{-\omega}= - i(y\partial+x_0 \partial_0 +x_1 \partial_1)
\partial\\&\qquad\qquad 
-\frac{1}{27}x_0 \partial_1^3
-\frac{4x_0}{x_1}\partial_1^2   -\frac{28x_0}{27x_1^2}\partial_1 
-6 i \partial\\&\qquad\qquad 
- \frac{x_1^3 \partial_0}{y^2} -i \frac{10x_1}{3y} \partial_1
-i\frac{x_1^2}{3y}\partial_1^2-6\frac{i}{y} \\ 
\end{array}
\nonumber
\end{equation}
\begin{equation}
H_\beta=-y \partial+x_0 \partial_0 \qquad 
H_\gamma = -y \partial-2x_0 \partial_0 -x_1 \partial_1 - 2 - 4s\, . 
\label{pioline:cube3}
\end{equation}
where $\partial \equiv \partial_y$ and $\partial_0\equiv \partial_{x_0}$. 
The virtue of this presentation is that the positive root 
Heisenberg algebra $[E_{\beta},E_{\gamma}]=E_{\omega}$ is canonically 
represented.
In addition, the Weyl reflection with respect to the root $\beta$ is
now very simple,
\begin{equation}
[A\cdot f](y,x_0,x_1) = e^{ i \frac{x_1^3}{x_0 y} }
f(-x_0,y,x_1)\, 
\label{pioline:Aflect}
\end{equation}
and the phase is cubic!
Notice that the same cubic term appears in the expression for
$E_{-\beta}$. Indeed, the spherical vector condition for the
compact generator $K_\beta=E_{\beta}+E_{-\beta}$ has solution
\begin{equation}
f_K(y,x_0,x_1)=\exp\Big[-\frac{ix_0x_1^3}{y(y^2+x_0^2)}\Big]
\,g(y^2+x_0^2)\, ,
\end{equation}
which implies an automorphic theta series formula summing 
over cubic rather than Gaussian characters~\cite{pioline:bump}. 

\section{Unipotent representations and theta series}
\index{Unitary representation!Unipotent}

The above construction of $Sl(3)$ 
Eisenstein series based on continuous series
representations extends easily to $Sl(n)$ and
(modulo some extra work) any simple Lie group:
they generalize the non-holomorphic $Sl(2)$ Eisenstein 
series~(\ref{pioline:eis2}).
However, the Jacobi theta series~(\ref{pioline:e1}) and its generalizations,
without any dependence on free parameters, is often more suited
to physical applications. Theta series can be obtained as residues 
of Eisenstein series at special points in their parameter space. 
Instead, here we wish to take a representation theoretic approach to 
theta series, based on automorphic forms coming from nilpotent orbits.

\subsection{The minimal representation of (A)DE groups} \index{ADE groups}
\index{Unitary representation!Minimal} 
The first step in gathering the various components of
formula~(\ref{pioline:eisd}) 
is to construct the {\it minimal} representation $\rho$ associated
to a nilpotent orbit of simple Lie groups $G$ other 
than $A_n$ (there are many different
constructions of the minimal representation in the literature, {\it e.g.}
\cite{pioline:KazhdanS,pioline:Kostant,pioline:Sahi,pioline:Gross,
pioline:Torasso},
see also~\cite{pioline:gunaydin} for a physicist's approach based
on Jordan algebras; we shall follow~\cite{pioline:KazhdanS}).
We will always consider the maximally split real form of $G$.
Minimality is ensured by selecting the nilpotent orbit of smallest
dimension: the orbit of the longest root 
$E_{-\omega}=j$ is a canonical choice. 
This orbit can be described by grading the Lie algebra ${\mathfrak g}$ 
with the Cartan generator $H_{\omega}=[E_{\omega},E_{-\omega}]$
(or equivalently studying the branching rule for the adjoint
representation under the $Sl(2)$ subgroup generated by $\{E_{\omega},H_{\omega},
E_{-\omega}\}$). The resulting 5-grading of  ${\mathfrak g}$ is
\begin{equation}
{\mathfrak g}={\mathfrak g}_{-2}\oplus{\mathfrak g}_{-1}\oplus{\mathfrak g}_{0}\oplus
{\mathfrak g}_{1}\oplus{\mathfrak g}_{2}
\end{equation}
where the one-dimensional 
spaces ${\mathfrak g}_{\pm 2}$ are spanned
by the highest and lowest roots $E_{\pm\omega}$.
Therefore the space ${\mathfrak g}_1 \oplus {\mathfrak g}_2$ 
is a Heisenberg algebra
of dimension ${\rm dim~ \mathfrak g}_1+1$ 
with central element $E_{\omega}$. 
Furthermore, since $[{\mathfrak g}_0,{\mathfrak g}_{\pm 2}]
={\mathfrak g}_{\pm 2}$, we have 
${\mathfrak g}_0 = {\mathfrak m} \oplus H_{\omega}$ where $[{\mathfrak m}, 
E_{\pm \omega}]=0$.
The Lie algebra ${\mathfrak m}$ generates the 
Levi subgroup $M$ of a parabolic group $P=MU$ 
with unipotent radical\footnote{Recall
that a parabolic group $P$ of upper block-triangular matrices (with
a  fixed given shape)
decomposes as $P=MU$ where the unipotent
radical $U$ is the subgroup
with unit matrices along the diagonal blocks 
while the Levi $M$ is the block diagonal  subgroup.}
$U=\exp {\mathfrak g}_{1}$. \index{Parabolic subgroup}
\index{Levi subgroup}
\index{Unipotent radical}
Hence 
the coadjoint orbit of $E_{-\omega}$ is parameterized by $H_{\omega}\oplus
{\mathfrak g}_{1} \oplus E_{\omega}$, the orthogonal
complement of its stabilizer. 
Its dimension is twice the dimension $d$ of the minimal representation
obtained through its quantization and is listed in Table~\ref{pioline:tlin}.

\begin{table}
$$\begin{array}
{c@{\hspace{4mm}}c@{\hspace{4mm}}c@{\hspace{4mm}}c@{\hspace{4mm}}c@{\hspace{4mm}}c}
G & d & M & L & {\mathfrak g}_1 & I_3 \\[.5mm]
\hline\hline
\\[-3mm]
Sl(n) & n-1 & Sl(n-2) & Sl(n-3) & \Real^{n-3} & 0\\
D_n & 2n-3 & Sl(2) \times D_{n-2} & D_{n-3} 
& \Real \oplus \Real^{2n-6} & x_1 (\sum x_{2i} x_{2i+1})\\ 
E_6 & 11 & Sl(6) & Sl(3)\times Sl(3) & \Real^3 \otimes \Real^3 & \det   \\
E_7 & 17 & SO(6,6) & Sl(6) & \Lambda^2\Real^6 & \mbox{Pf} \\
E_8 & 29 & E_7 & E_6 & 27 & 27^{\otimes_s 3}\vert_{1}
\end{array}$$
\caption{Dimension of minimal representations, 
canonically realized Levi subgroup~$M$,
linearly realized subgroup $L$,
representation of~${\mathfrak g}_1$ under~$L$ and
cubic~$L$-invariant~$I_3$.
\label{pioline:tlin}}
\end{table}

To quantize the minimal nilpotent orbit,  
note that the symplectic vector space ${\mathfrak g}_{1}$ admits a canonical
polarization chosen by taking as momentum variables the positive
root $\beta_0$ attached to the  highest root $\omega$ on the extended
Dynkin diagram, along with those positive roots $\beta_{i=1,\dots, d-2}$
with Killing inner products $(\beta_0 , \beta_i)=1$.
The conjugate position variables are then $\gamma_{i=0,\dots, d-2}=
\omega-\beta_i$. These generators are given by the Heisenberg representation 
$\rho_H$ acting on functions
of $d$ variables,
\begin{equation}
\label{pioline:heis}
E_\omega=iy\ , \qquad E_{\beta_i}=y\ \partial_{x_0}\ ,\qquad E_{\gamma_i}=ix_0\ ,
\quad i=0,\dots, d-2\, .
\end{equation}
So far the generator $y$ is central. By the 
Shale--Weil theorem~\cite{pioline:Weil},\index{Shale--Weil theorem}
$\rho_H$ extends to a representation of the
double cover of the symplectic group $Sp(d-1)$. The latter contains 
the Levi $M$ with trivial central extension of $Sp(2d)$ over~$M$. 
In physics terms, the Levi $M$ acts linearly on the positions
and momenta by canonical transformations. In particular,
the longest element $S$ in the Weyl group of $M$ is represented by 
Fourier transform,
\begin{equation}
\label{pioline:eS}
[S \cdot f](y,x_0,\dots,x_{d-2})=\int 
\left[\prod_{i=0}^{d-2}\frac{dp_i}{\sqrt{2\pi y}}\right]
\, f(y,p_0,\dots,p_{d-2})\,\  e^{\frac{i}{y}\sum_{i=0}^{d-2}p_i x_i}\, .
\end{equation}
The subgroup $L\subset M$ commuting with $E_{\beta_0}$, 
does not mix positions and momenta and therefore
acts linearly on the variables $x_{i=1\dots d-2}$
while leaving $(y,x_0)$ invariant. The representation
of the parabolic subgroup $P$ can be extended to  
$P_0=P\times \exp{t H_{\beta_0}}$ (where $\exp{t H_{\beta_0}}$ is the 
one-parameter subgroup generated by $H_{\beta_0}=[E_{\beta_0},
E_{-\beta_0}]$) by defining
\begin{equation}
H_{\beta_0} = -y \partial + x_0 \partial_{0}\, ,
\end{equation}
(here $\partial\equiv \partial_y$ 
and $\partial_i\equiv\partial_{x_i}$). Notice that the
element $y$, which played the r\^ole of $\hbar$ before, is no longer
central.
To extend this representation to the whole of $G$, note
that Weyl reflection with respect to the root $\beta_0$ acts just 
as in the $Sl(3)$ case~(\ref{pioline:Aflect}),
\begin{equation}
\label{pioline:eA}
[A\cdot f](y,x_0,x_1,\dots,x_{d-2})
= e^{-\frac{iI_3}{x_0 y}} f(-x_0,y,x_1,\dots,x_{d-2})\, .
\end{equation}
In this formula, $I_3(x_i)$ is the unique $L$-invariant 
(normalized) homogeneous, cubic, polynomial in the $x_{i=1,\dots,
d-2}$ (see Table~\ref{pioline:tlin}).
Remarkably, the Weyl group relation
\begin{equation}
\label{pioline:e241}
(AS)^3=(SA)^3
\end{equation}
holds, thanks to the invariance of the cubic character
$e^{-iI_3/x_0}$ under Fourier transform over 
$x_{i=0\dots d-2}$~\cite{pioline:etingof} (see also~\cite{pioline:Pioline:2003uk}).
This is the analog of the Fourier invariance of the Gaussian character
for the symplectic theta series. It underlies
the minimal nilpotent representation and its 
theta series.

The remaining generators are obtained by
applying the Weyl reflections $A$ and $S$ to the Heisenberg
subalgebra~(\ref{pioline:heis}). 
In particular, the negative root $E_{-\beta_0}$
takes the universal form,
\begin{equation}
E_{-\beta_{0}} =- x_{0} \partial + \frac{i I_3}{y^2}
\end{equation}
which we first encountered in the $Sl(3)$ example~(\ref{pioline:cube3}).

It is useful to note that this construction can be cast in the language
of Jordan algebras: $L$ is in fact the reduced structure group of a 
cubic Jordan algebra $J$ with norm $I_3$; $M$ and $G$ can then be understood
as the ``conformal'' and ``quasi-conformal'' groups associated to $J$.
\index{Quasi-conformal}
The minimal representation arises from quantizing the quasi-conformal
action -- see~\cite{pioline:Gunaydin:2000xr} for more details on this
approach, which generalizes to all semi-simple algebras including
the non simply-laced cases.

\subsection{$D_4$ minimal representation and strings on $T^4$}

As illustration, we display the minimal representation 
of $SO(4,4)$~\cite{pioline:Kazhdan} (see~\cite{pioline:Kostant0}
for an alternative construction). The extended Dynkin diagram is
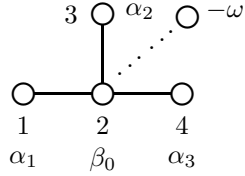
\index{$SO(4,4,\Zint)$}
\begin{center}
\begin{picture}(100,35)
\thicklines
\multiput(0,0)(30,0){3}{\circle{8}}
\put(0,-12){\makebox(0,0){1}}\put(0,-25){\makebox(0,0){$\alpha_1$}}
\put(30,-12){\makebox(0,0){2}}\put(30,-25){\makebox(0,0){$\beta_0$}}
\put(60,-12){\makebox(0,0){4}}\put(60,-25){\makebox(0,0){$\alpha_3$}}
\multiput(4,0)(30,0){2}{\line(1,0){22}}
\put(30,4){\line(0,1){22}}
\put(30,30){\circle{8}}
\put(18,30){\makebox(0,0){3}}\put(44,31){\makebox(0,0){$\alpha_2$}}
\put(34,6){$\adots$}
\put(46,16){$\adots$}
\put(69,29){$-\omega$}
\put(62,29){\circle{8}}
\end{picture}
\end{center}
\vskip 10mm
\noindent 
and the affine root $-\omega$ attaches to the root $\beta_0$. 
The grade-1 symplectic vector space~${\mathfrak g}_1$ is spanned by
$4+4$ roots
\begin{equation}
\left.
\begin{array}{ccc}
\beta_0&\quad&\gamma_0=\beta_0+\alpha_1+\alpha_2+\alpha_3\\
\beta_i=\beta_0+\alpha_i&&\gamma_i=\beta_0+\alpha_j+\alpha_k
\end{array}
\right\}\quad \{i,j,k\}=\{1,2,3\}\, .
\end{equation}
The positive roots are represented as in~(\ref{pioline:heis}), while the
negative roots read
\begin{eqnarray}
\label{pioline:cube4}
E_{-\beta_{0}}&=&- x_{0} \partial + \frac{i x_{1} x_{2} x_{3}}{y^2}\,  \nonumber\\
E_{-\beta_{1}}&=& x_{1} \partial + \frac{x_{1}}{y}\,  ( 1+ x_{2} \partial_{2}
+ x_{3} \partial_{3}) - i x_{0} \partial_{2} \partial_{3} \nonumber\\
\label{pioline:e30}
E_{-\gamma_{0}}&=&3i\partial_0+iy\partial\partial_0
-y\partial_1\partial_2\partial_3
         +i(x_0\partial_0+x_1\partial_1+x_2\partial_2+x_3\partial_3)\, 
\partial_0 \nonumber\\
E_{-\gamma_{1}}&=& i y \partial_{1} \partial  + i(2+ x_{0} \partial_{0}
+ x_{1} \partial_{1})\,  \partial_{1}  
- \frac{x_{2} x_{3}}{y}\,  \partial_{0}\nonumber\\
E_{-\omega}&=&3i\partial + iy\partial^2 +\frac{i}{y} 
+ ix_0 \partial_0\partial+ 
\frac{x_1x_2x_3}{y^2}\, \partial_0  +x_0\partial_1\partial_2\partial_3 \nonumber \\
&& + \frac{i}{y}\, (x_1x_2\partial_1\partial_2+ \mbox{cyclic} )
+ \ i(x_1\partial_1+x_2\partial_2+x_3\partial_3)\, (\partial+\frac{1}{y})\, ,
\end{eqnarray}
as well as cyclic permutations of $\{1,2,3\}$.
The Levi $M=[Sl(2)]^3$, obtained by removing $\beta_0$ from the
extended Dynkin
diagram, acts linearly on positions and momenta and has generators
\begin{equation}
E_{\alpha_i}=- x_{0} \partial_{i} -\frac{i  x_{j} x_{k}}{y} \ ,\quad
E_{-\alpha_i}= x_{i} \partial_{0} +i  y \partial_{j} \partial_{k} \, .
\end{equation}
Finally, the Cartan generators are obtained from 
commutators $[E_{\alpha},E_{-\alpha}]$,
\begin{eqnarray}
H_{\beta_0}&=&- y \partial + x_{0} \partial_{0}\\
H_{\alpha_i}&=&\phantom{- y \partial} 
- x_{0} \partial_{0} + x_{i} \partial_{i} 
- x_{j} \partial_{j} - x_{k} \partial_{k}-1\label{pioline:e31}\, .
\end{eqnarray}

This representation also arises in a totally different context: 
the one-loop amplitude for 
closed strings compactified on a 4-torus! 
$T$-duality requires this amplitude to
be an automorphic form of $SO(4,4,\Zint)=D_4(\Zint)$. \index{T-duality} 
In fact, it may be written as an integral of
a symplectic theta series over the fundamental domain of
the genus-1 world-sheet moduli space,
\begin{equation}
\label{pioline:1loops}
{\cal A}(g_{ij},B_{ij})=\int_{SO(2)\backslash Sl(2,\Real)
/Sl(2,\Zint)} \frac{d^2\tau}{\tau_2^2}\,\, 
\theta_{Sp(8,\Zint)}(\tau,\bar\tau;g_{ij},B_{ij})\, .
\end{equation}
Here $(g_{ij},B_{ij})$ are the metric and Neveu-Schwarz two-form
on $T^4$ parameterizing the moduli space $[SO(4)\times SO(4)]\backslash
SO(4,4,\Real)$. The symplectic theta series 
$\theta_{Sp(8,\Zint)}$ is the partition function of the
$4+4$ string world-sheet winding modes $m^i_a$, $i=1,\dots, 4, a=1, 2$  
around $T^4$.  Like any Gaussian theta series, it is 
invariant under the (double cover of the) 
symplectic group over integers, $Sp(8,\Zint)$ in this case. 
The modular group and T-duality group arise as a dual pair
$Sl(2) \times SO(4,4)$ in $Sp(8)$ -- in other words, each factor 
is the commutant of the other within $Sp(8)$.
Therefore, after integrating over the $Sl(2)$ moduli space,
an $SO(4,4,\Zint)$ automorphic form, based on the minimal representation 
remains. Dual pairs are a powerful technique for constructing
new automorphic forms from old ones\index{Dual pairs}.

To see the minimal representation of $D_4$ emerge explicitly, 
note that $Sl(2)$-invariant 
functions of $m^i_a$ must depend on the cross
products, 
\begin{equation}
\label{pioline:sl4c}
m^{ij}=\epsilon^{ab} m^i_a m^j_b \ ,
\end{equation}
which obey the quadratic constraint
\begin{equation} 
m^{[ij} m^{kl]} = 0\, , 
\end{equation}
and therefore span a cone in $\Real^6$.
The 5 variables $(y,x_0,x_i)$ are mapped to this 
5-dimensional cone by diagonalizing the action of the 
maximal commuting set of six observables 
${\cal C}=(E_{\alpha_3},E_{\beta_3},E_{\gamma_1},E_{\gamma_2},E_{\gamma_0},
E_\omega)$
whose eigenvalues may be identified with 
the constrained set of six coordinates on the cone
$i(m^{43},m^{24},m^{14},$ $m^{23},m^{13},m^{12})$.
The intertwiner between
the two representations is a convolution with the common eigenvector
of the generators ${\cal C}$  amounting to a Fourier transform over $x_3$.
This intertwiner makes the hidden triality symmetry,
which is crucial for heterotic/type II duality~\cite{pioline:Kiritsis:2000zi}, \index{Superstring dualities} \index{Triality}
of the $Sl(4)$-covariant 
string representation manifest.

An advantage of the covariant realization is that the
spherical vector follows by directly computing 
the integral~(\ref{pioline:1loops}). The real
spherical vector is read-off from 
worldsheet instanton contributions \index{Worldsheet instantons}
\begin{equation}
\label{pioline:e39}
f_{\infty}(m^{ij})=\frac{e^{-2\pi \sqrt{(m^{ij})^2}}}{\sqrt{(m^{ij})^2}}\, ,
\end{equation}
while its $p$-adic counterpart follows from the
instanton summation measure
\begin{equation}
\label{pioline:e39p}
f_p(m^{ij})=\gamma_p(m_{ij}) \frac{1-p \,|m_{ij}|_p}{1-p}\, .
\end{equation}
Intertwining back to the triality invariant realization gives
\begin{equation}
\label{pioline:e49}
f_{\infty}(y,x_0,x_i)=\frac{e^{ - i\, \frac{x_0 x_1 x_2 x_3}{y (y^2 + x_0^2)}}}
{\sqrt{y^2+x_0^2}}\, K_0 \left( 
\frac{\sqrt{\prod_{i=1}^3(y^2+x_0^2+x_i^2)}}
{y^2+x_0^2} \right)\, .
\end{equation}
This is the prototype for spherical vectors of all higher simple Lie groups.

\subsection{Spherical vector, real and $p$-adic} \index{Spherical vector!real} \index{Spherical vector!$p$-adic}

To find the spherical vector for higher groups, one may either
search for generalizations of the covariant string representation
in which the result is a simple extension of the world-sheet
instanton formula~(\ref{pioline:e39}) -- see Section~\ref{pioline:pure},
or try and solve by brute force the complicated set of partial differential
equations $(E_{\alpha}+E_{-\alpha})f=0$ demanded by $K$-invariance.
Fortunately, knowing the exact solution (\ref{pioline:e49}) for $D_4$
gives enough inspiration to solve the general case~\cite{pioline:kpw}.

To see this, note that the phase in \eqref{pioline:e49}
has precisely the right anomalous transformation under
$(y,x_0)\to(-x_0,y)$ to cancel the cubic character 
of the Weyl generator~(\ref{pioline:eA}), or equivalently
the cubic term appearing in $E_{\beta_0}+E_{-\beta_0}$.
The real part of the spherical vector therefore 
depends on $(y,x_0)$ through their norm $R=\sqrt{y^2+x_0^2}=|y,x_0|_\infty$.
Moreover, invariance under the linearly acting maximal compact subgroup 
of~$L$ restricts the dependence on $x_i$ to its quadratic
$I_2$, cubic $I_3$ and quartic $I_4$ invariants. Choosing a frame where
all but three of the $x_i$ vanish, the remaining equations
are then essentially the same as for the known $D_4$ case.
The universal result is
\begin{equation}
\label{pioline:esph}
f_{\infty}(X)=\frac{1}{R^{s+1}}\,
{\cal K}_{s/2}\left( \left| X,\nabla_X [ I_3/R] \right|_{\infty} \right)\, 
\exp\left(-i\, \frac{x_0I_3}{yR^2}\right)\, ,
\end{equation}
where $X\equiv(y,x_0,\dots x_{d-2})$ and $I_3$ is given in Table~\ref{pioline:tlin}.
Notice that the result depends on the pullback of the
Euclidean norm to the Lagrangian subspace
$(X,\nabla_X[I_3/R])$ of the coadjoint orbit.
The function 
${\cal K}_t(x)$ is related to the
usual modified Bessel function by ${\cal K}_t(x)\equiv x^{-t}K_t(x)$,
and the parameter $s=0,1,2,4$ for $G=D_4, E_6, E_7, E_8$, 
respectively\footnote{For $D_{n>4}$ the result is slightly more
complicated, see~\cite{pioline:kpw}. It is noteworthy that
the ratio $I_3/R$ is invariant under Legendre transform with respect to
all entries in $X$, although the precise meaning of this observation
is unclear.}.

The $p$-adic spherical vector computation is much harder since
the generators cannot be expressed as 
differential operators. It was nevertheless
completed in~\cite{pioline:sasha} by very different techniques, again
inspired by the $D_4$ result~(\ref{pioline:e39p}), intertwined to the
triality invariant representation.
The result mirrors the real case, 
namely for $|y|_p<|x_0|_p$,
\begin{equation}
\label{pioline:esphp}
f_{p}(X)=\frac{1}{R^{s+1}}\,
{\cal K}_{p,s/2}\left( X,\nabla_X [ I_3/R] \right)\, 
\exp\left(-i\, \frac{I_3}{x_0 y}\right)\, ,
\end{equation}
where $R=|y,x_0|_p=|x_0|_p$ is now the $p$-adic norm, and 
${\cal K}_{p,t}$ is a $p$-adic analogue of the modified Bessel function,
\begin{equation}
{\cal K}_{p,t}(x) = \frac{1 - p^s |x|_p^{-s} } {1- p^s}\  \gamma_p(x)\, ,
\end{equation}
($\gamma_p(x)$ generalizes to a function of
several arguments by $\gamma_p(X)=0$ unless $|X|_p\leq 1$). 
The result for $|y|_p>|x_0|_p$ follows by the
Weyl reflection $A$.

\subsection{Global theta series}
Having obtained the real and $p$-adic spherical vectors for any $p$,
one may now insert them in the adelic formula~(\ref{pioline:adelsum}) to
construct exceptional theta series. Equivalently,
we may use the representation~(\ref{pioline:eisd}), \index{Theta series!adelic representation}
\begin{equation}
\theta_G(e)=\langle \delta_{G(\Zint)}, \rho_G(e) f_{\infty}\rangle\ ,
\qquad \delta(X) = \prod_{p\ {\rm prime}} f_{p}(X)\, .
\end{equation}
Thanks to the factor $\gamma_p(X,
\nabla_X[I_3/R])$, the summation measure $\delta_{G(\Zint)}(X)$ 
will have support
on integers $X$ such that $\nabla_X[I_3/R]$ is also an integer.

While this expression is fine for generic $X$, it ceases to make
sense when $y=0$, as the phase of the spherical vector~(\ref{pioline:esph})
becomes singular. As shown in~\cite{pioline:sasha}, the correct prescription
for $y=0$ is to remove the phase and set $y=0$ in the rest of the 
spherical vector, thereby obtaining a new smooth vector 
\begin{equation}
\overline f(\overline X) = \lim_{y\to 0} \left[
\exp\left(i\, \frac{x_0 I_3}{yR^2}\right) f_{\infty}(y,x_0,x_i) \right]\, ,
\end{equation}
where $\overline X=(x_0,x_1,\dots x_{d-2})$,
with a similar expression in the $p$-adic case. 

However, there still remains a further divergence when $y=x_0=0$.
It can be shown that these terms may be regularized to give a
sum of two terms, namely a constant plus a theta series based
on the minimal representation of the Levi subgroup $M$. Altogether,
the global formula for the theta series in the minimal representation
of $G$ reads~\cite{pioline:sasha}
\begin{eqnarray}
\theta_G(e) &=& \sum_{X\in [\Zint\backslash \{0\}] \times \Zint^{d-1} }
\mu(X) ~ \rho(e)\cdot f_{\infty}(X) \nonumber\\
&&+ \sum_{\overline X\in [\Zint\backslash \{0\}] \times \Zint^{d-2}} 
\overline\mu(\overline X) ~ \rho(e)\cdot \overline f_{\infty}(\overline X) 
+ \alpha_1 + \alpha_2 \theta_M(e) \, .
\end{eqnarray}
Notice that the degenerate contributions in the second line will mix
with the non-degenerate ones under a general right action of $G(\Zint)$.

\subsection{Pure spinors, tensors, 27-sors, \dots} \index{Pure spinors}
\label{pioline:pure}

We end the mathematical discussion by returning to
the $Sl(4)$-covariant presentation of the minimal representation 
of $SO(4,4)$ on functions of 6 variables $m^{ij}$ with a quadratic 
constraint~(\ref{pioline:sl4c}). The existence of this
presentation may be traced to the 3-grading $28 = 
6 \oplus (15+1) \oplus 6$ of the Lie algebra of $SO(4,4)$
under the Abelian factor in $Gl(4)\subset SO(4,4)$: the top space 
in this decomposition is an Abelian group, whose generators in the
minimal representation of $SO(4,4)$ can be simultaneously diagonalized.
The eigenvalues transform linearly under $Sl(4)$ as a
two-form, but satisfy one constraint in accord with the functional
dimension 5 of the minimal representation of $SO(4,4)$.

This phenomenon also occurs for higher groups: for $D_n$, the 
branching of $SO(n,n)$ into $Gl(n)$ leads to a dimension $n(n-1)/2$
Abelian subgroup, whose generators transform linearly
as antisymmetric $n\times n$  matrices~$m^{ij}$. Their
simultaneous diagonalization in the minimal representation of $D_n$
leads to the same constraints as in~(\ref{pioline:sl4c}), 
solved by rank 2 matrices~$m^{ij}$. The number of independent variables
is thus $2n-3$, in accord with the functional
dimension of the minimal representation. This is in fact the presentation
obtained from the dual pair $SO(n,n)\times Sl(2) \subset Sp(2n)$,
and just as in~(\ref{pioline:e39}), the spherical vector is a Bessel 
function of the norm $\sqrt{(m^{ij})^2}$.

For $E_6$, the 3-grading $78 = 16 \oplus (45+1) \oplus 16$ from
the branching into $SO(5,5)\times \Real$ leads to a realization
of the minimal representation of $E_6$ on a spinor $Y$ of $SO(5,5)$,
with 5 quadratic constraints $\overline Y \Gamma_{\mu} Y=0$. 
The solutions to these constraints
are in fact the pure spinors of Cartan and Chevalley\index{Pure spinors}. 
The spherical vector was computed in~\cite{pioline:kpw} by Fourier
transforming over one column of the $3\times 3$ matrix $X$ appearing 
in the canonical polarization, and 
takes the remarkably simple form
\begin{equation}
f_{\infty}(Y) = {\cal K}_1 \left(\sqrt{ \overline Y Y } \right)\, .
\end{equation}
Its $p$-adic counterpart, obtained by replacing orthogonal 
with $p$-adic norms, also simplifies accordingly.
We thus conclude
that functions of pure spinors of $SO(5,5)$ (as well as other real forms of 
$D_5$) carry an action of $E_{6(6)}(\Real)$\footnote{In 
contrast to the conformal realization of $E_6$ on 21 variables discussed 
in~\cite{pioline:Gunaydin:2000xr}, this  representation is irreducible.}. 
Given that pure spinors of $SO(9,1)$ provide a convenient 
covariant reformulation
of ten-dimensional super-Yang-Mills theory 
and string theory~\cite{pioline:Berkovits:2000fe}, 
it is interesting to ponder the physical consequences of this hidden 
$E_6$ symmetry.

For $E_7$, the 3-grading corresponding to
the branching $133=27\oplus (78+1) \oplus 27$ into $E_6 \times \Real$,
leads to a realization of the minimal representation of $E_7$ on
a 27 representation of $E_6$, denoted $Y$,
subject to the condition that the $\overline{27}$ part in the 
symmetric tensor product $27 \otimes_s 27$ vanishes -- in
other words, $\partial_Y I_3(Y)=0$. This corresponds to 10 independent
quadratic conditions, whose solutions may aptly be dubbed {\it pure 27-sors}. 
The spherical vector was computed in~\cite{pioline:kpw} by
Fourier transforming over one column of the 
antisymmetric $6\times 6$ matrix $X$ in the canonical polarization,
and is again extremely simple
\begin{equation}
f_{\infty}(Y) = {\cal K}_{3/2} \left(\sqrt{ \overline Y Y } \right)\, .
\end{equation}

Unfortunately, $E_8$ does not admit any 3-grading. However,
the 5-grading $248 = 1\otimes 56 \otimes (133+1) \otimes 56 \otimes 1$
from the branching into $E_7 \times Sl(2)$ leads to an action of $E_8$
on functions of ``pure'' 56-sors $Y$ of $E_7$
together with an extra variable $y$. For the minimal representation
of $E_8$, the appropriate notion of purity requires the quadratic equations
$\partial_Y \otimes \partial_Y I_4(Y)=0$, where $I_4$ is the quartic
invariant of $E_7$. As explained in~\cite{pioline:Gross}, less stringent
purity conditions lead to unipotent representations with larger dimension.
This kind of construction based on a 5-grading is in fact available 
for all semi-simple groups in the quaternionic real form, and is 
equivalent to the ``canonical'' 
construction of the minimal representation  in the simply-laced 
case~\cite{pioline:Gross}.

\section{Physical applications}
Having completed our brief journey into the dense forest
of unipotent representations and automorphic forms, we
now return to a more familiar ground, and describe
some physical applications of these mathematical constructions.
 
\subsection{The automorphic membrane}
\index{Supermembrane}
\index{M-theory}

The primary motivation behind our study of
exceptional theta series was the conjecture of~\cite{pioline:pnpw}:
the exact four-graviton~$R^4$ scattering amplitude, predicted
by $U$-duality and supersymmetry, \index{Superstring dualities} 
ought be derivable 
from the eleven-dimensional quantum supermembrane -- an obvious
candidate to describe fundamental $M$-theory excitations.
For example, in eight dimensions, 
in analogy with the one-loop string amplitude, 
the partition function of supermembrane
zero-modes should be a theta series of~$E_6(\Zint)$, which
subsumes both the $U$-duality group $Sl(3,\Zint)
\times Sl(2,\Zint)$, and the toroidal 
membrane modular group $Sl(3,\Zint)$. Integrating the partition function
over world-volume moduli $\Real^+ \times Sl(3)$, 
yields by construction a $U$-duality invariant result 
which should reproduce the
exact four-graviton $R^4$ scattering amplitude for M-theory on a~$T^3$, 
including membrane instantons, 
namely a sum of $Sl(3)$ and $Sl(2)$ Eisenstein 
series~\cite{pioline:Kiritsis:1997em,pioline:Obers:1999um}.

Having constructed explicitly the $E_6$ theta series, we may now
test this conjecture~\cite{pioline:pw03}. Recall that in the canonical realization, 
the $E_6$ minimal representation contains an $Sl(3) \times
Sl(3)$ group acting linearly from the left and  right
on a $3\times 3$ matrix of integers $m_{M}^A$, together with two
singlets $y,x_0$. In addition, there is an extra $Sl(3)$
built from  the non-linearly acting generators $E_{\beta_0,\gamma_0,\omega}$,
which further decomposes into the $\Real^+ \times Sl(2)$ factors mentioned
above. The integers  $m_{M}^A$ are interpreted as
winding numbers of a toroidal membrane wrapping the target-space $T^3$,
$X^M= m^M_A \sigma^A$. The two extra integers $y,x_0$ do
not appear in the standard membrane action but may
be interpreted as a pair of world-volume 3-form fluxes -- an
interesting prediction of the hidden $E_6$ symmetry, recently
confirmed from very different arguments~\cite{pioline:Bengtsson}.

The integration over the membrane world-volume $Sl(3)$ 
moduli 
amounts to decomposing the minimal representation with respect to
the left acting $Sl(3)$ and keeping only invariant singlets. 
For a generic matrix $m_{M}^A$,
the unique such invariant is its determinant, which we preemptively
denote $x_1^3 = \det(M)$. This leaves a representation of
the non-linear $Sl(3)$ acting
on functions of three variables $(y,x_0,x_1)$ 
(the right $Sl(3)$ acts trivially): this is precisely the
representation studied in Section~\ref{pioline:gafgg}. In addition, non-generic 
matrices contribute further representations charged under 
both left and right $Sl(3)$s.

It remains to carry out the integration over the membrane world-volume
factor $\Real^+$ inside the non-linear $Sl(3)$. This integral
is potentially divergent. 
Instead, a correct
mathematical prescription is to look at the constant term with respect
to a parabolic $P_{1,2} \subset Sl(3)_{NL}$: indeed we find that
this produces the result predicted by the conjecture~\cite{pioline:pw03}.

This is strong evidence 
that  membranes are indeed the correct 
degrees of freedom of M-theory, although the construction only
treats membrane zero-modes. It would be very 
interesting to see if the $E_6$ symmetry can be extended to 
fluctuations and in turn to lead to a quantization of the
complete toroidal supermembrane.

\subsection{Conformal quantum cosmology} \index{Cosmology}

The dynamics of spatially separated points decouple
as a space-like singularity is approached.
Only effective 0+1-dimensional
quantum mechanical degrees of freedom remain 
at each point. Classically, these 
correspond to a particle on a
hyperbolic billiard, whose chaotic motion translates into a sequence 
of Kasner flights and bounces of the spatial geometry~\cite{pioline:bkl}.
Originally observed for 3+1-dimensional Einstein
gravity, this chaotic behavior persists for 11-dimensional
supergravity, whose the billiard is the
Weyl chamber of a generalized $E_{10}$ Kac-Moody group~\cite{pioline:Damour:2000wm}. 
Upon accounting for off-diagonal metric and gauge 
degrees of freedom, the hyperbolic billiard can be unfolded 
onto the fundamental domain of the arithmetic group $E_{10}(\Zint)$. 
\index{$E_{10}({\mathbb Z})$}
Automorphic forms for $E_{10}(\Zint)$ should therefore be relevant in
to the wave function of the universe!\index{Wave function of the universe} 

Automorphic forms for generalized Kac-Moody groups
are out of our present reach. However,
automorphic forms for finite Lie groups may still be useful 
in a cosmological context because their
corresponding minimal representations can be viewed as conformal
quantum mechanical systems of the type that arising near cosmological
singularities~\cite{pioline:Pioline:2002qz}. Indeed, changing variables 
$y=\rho^2/2$, $x_i = \rho q_i/2$ in the canonical minimal representation,
the generators of the grading $Sl(2)$ subalgebra become
\begin{equation}
\label{pioline:so212}
{E_{\omega}}=\frac12\rho^2\ ,\quad
H_{\omega}=\rho p_\rho\ ,
\quad E_{-\omega}=\frac12 \left( p_\rho^2 + 
\frac{4\Delta}{\rho^2} \right)\, .
\end{equation}
Here $\Delta$ is
a quartic invariant of the coordinates and momenta 
$\{q_i,\pi_i\}$ corresponding (up to an additive constant) to
the quadratic Casimir of the Levi~$M$.
Choosing~$E_{-\omega}$ as the Hamiltonian, 
the resulting mechanical system has a dynamical, $d=0+1$ conformal, 
$Sl(2)=SO(2,1)$ symmetry.
In contrast to the one-dimensional conformal quantum mechanics 
of~\cite{pioline:fubini}, the conformal symmetry~$Sl(2)$ is
enlarged to a much larger group~$G$ mixing the radial coordinate~$\rho$ 
with internal ones~$x_i$. It can be shown that these
conformal systems appear upon  dimensional reduction of Einstein's 
equations near a space-like singularity~\cite{pioline:Pioline:2002qz}. 

\subsection{Black hole micro-states} \index{Black holes}
Finally, minimal representations and automorphic forms 
play an important r\^ole in understanding the microscopic
origin of the Bekenstein-Hawking entropy of black holes. 
{}From thermodynamic arguments, these stationary, spherically symmetric 
classical solutions of Einstein-Maxwell gravity are expected to describe
an exponentially large number of quantum micro-states 
(on the order of the exponential 
of the area of their horizon in Planck units). 
It is an important question to determine
the exact degeneracy of micro-states for a given value of their charges --
as always, U-duality is a powerful constraint on the result.
An early conjecture in the framework of $N=4$ supergravity relates
the degeneracies to Fourier coefficient of a certain modular form
of $Sp(4,\Zint)$ constructed by Igusa~\cite{pioline:dvv}. A~more recent
study suggests that the 3-dimensional U-duality group (manifest
after timelike dimensional reduction of the 4-dimensional stationary 
solution) should play the r\^ole of a ``spectrum generating
symmetry'' for the black hole degeneracies~\cite{pioline:pio2005}. 
For M-theory compactified
on $T^7$ or $K3\times T^3$, the respective
$E_8(\Zint)$ or $SO(8,24,\Zint)$ symmetry may be sufficiently powerful
to determine these degeneracies, and there are strong indications that
the minimal representation and theta series are the appropriate 
objects~\cite{pioline:pio2005,pioline:Gunaydin:2005gd,pioline:gnpw}.

\section{Conclusion}

In this Lecture, we hope to have given a self-contained introduction
to automorphic forms, based on string theory experience -- rigor
was jettisoned in favor of simplicity and utility.
Our attempt will be rewarded if the reader is
preempted to study further aspects of this rich field: 
non-minimal unipotent representations, non-simply laced 
groups, non-split real forms, reductive dual pairs,
arithmetic subgroups, Fourier coefficients, L-functions...
Alternatively, he or she may solve any of the homework problems 
outlined in Section 4. 

\vskip 2mm
\noindent {\it Acknowledgments:}~ We would like to thank our mathematical muse
D. Kazhdan and his colleagues S. Miller, C. Moeglin, S. Polischchuk
for educating us and the Max Planck Institute f\"ur Mathematik Bonn and 
Gravitationsphysik -- Albert Einstein Institut -- in Golm for hospitality
during part of this work. B.P. is also grateful for the organizers of Les Houches Winter School on ``Frontiers in Number Theory, Physics 
and Geometry'' for a wonderful session, and the kind invitation to 
present this work. Research 
supported in part by NSF grant PHY01-40365.

%
%
%

\printindex
\end{document}